\documentclass[final]{IEEEtran}
\pdfoutput=1
\IEEEoverridecommandlockouts

\usepackage{cite}
\usepackage{amsmath,amssymb,amsfonts, amsthm}
\usepackage{gensymb}
\usepackage{algorithmic}
\usepackage{graphicx}
\usepackage{float}
\usepackage{subcaption}
\usepackage{textcomp}
\usepackage{xcolor}
\usepackage{hyperref}

\def\BibTeX{{\rm B\kern-.05em{\sc i\kern-.025em b}\kern-.08em
    T\kern-.1667em\lower.7ex\hbox{E}\kern-.125emX}}
\begin{document}

\title{Aircraft Radar Altimeter Interference Mitigation Through a CNN-Layer Only Denoising Autoencoder Architecture}

\author{\IEEEauthorblockN{Samuel B. Brown\IEEEauthorrefmark{1}, Stephen Young\IEEEauthorrefmark{2}, Adam Wagenknecht\IEEEauthorrefmark{2}, Daniel Jakubisin\IEEEauthorrefmark{1}, \\Charles E. Thornton\IEEEauthorrefmark{1}, Aaron Orndorff\IEEEauthorrefmark{1}, and William C. Headley\IEEEauthorrefmark{1}}
\IEEEauthorrefmark{1}Virginia Tech National Security Institute,
Blacksburg, VA, USA \\
\IEEEauthorrefmark{2}The Boeing Company, Crystal City, VA, USA}

\maketitle

\thispagestyle{plain}
\pagestyle{plain}

\maketitle

\begin{abstract}
Denoising autoencoders for signal processing applications have been shown to experience significant difficulty in learning to reconstruct radio frequency communication signals, particularly in the large sample regime. In communication systems, this challenge is primarily due to the need to reconstruct the modulated data stream which is generally highly stochastic in nature. In this work, we take advantage of this limitation by using the denoising autoencoder to instead remove interfering radio frequency communication signals while reconstructing highly structured FMCW radar signals. More specifically, in this work we show that a CNN-layer only autoencoder architecture can be utilized to improve the accuracy of a radar altimeter's ranging estimate even in severe interference environments consisting of a multitude of interference signals. This is demonstrated through comprehensive performance analysis of an end-to-end FMCW radar altimeter simulation with and without the convolutional layer-only autoencoder. The proposed approach significantly improves interference mitigation in the presence of both narrow-band tone interference as well as wideband QPSK interference in terms of range RMS error, number of false altitude reports, and the peak-to-sidelobe ratio of the resulting range profile. FMCW radar signals of up to 40,000 IQ samples can be reliably reconstructed. 
\end{abstract}


\section{Introduction}
Due to the increased proliferation of sub-6~GHz 5G NR signals in recent years, it has been shown that there is potential of increased interference on radar altimeter systems. Such unwanted interference could cause errors in ground range estimates, particularly during aircraft landing \cite{remcom, rtca, Griffiths2015}. Moreover, the rapid increase of automotive radar systems, generally operating in the same $77$GHz frequency band, has drawn much attention to interference mitigation techniques \cite{Uysal2018}. In both of these application areas, the radar systems utilize frequency modulated continuous wave (FMCW) signals along with stretch processing techniques for range estimation. 

While a variety of signal processing techniques have been proposed for interference mitigation \cite{Aydogdu2020,oyedare2022interference}, the application of \emph{denoising autoencoders} to remove interference is particularly compelling for several reasons. First, no initial detection or estimation of the interfering signal is needed. Second, no explicit assumptions are made regarding the interference and noise distributions. Finally, with proper training, the model may generalize across a variety of interference types. Denoising autoencoders utilize a latent space with lower dimensionality than the input/output space to promote generalization of important data features and discarding of unwanted information such as noise and interference \cite{Kokalj2019, Fuchs2020, Chen2021,oyedare2022interference}.

However, denoising autoencoders are accompanied by a number of practical challenges. First, and most importantly, there is a natural trade-off between the reconstruction capabilities of the autoencoder and its ability to sufficiently compress the latent space for successful denoising. For communication signals, the natural stochastic nature of the signal structure (particularly the modulated data stream) is extremely difficult to compress into a latent space making denoising of interference signals particularly challenging. Secondly, for the particular case of high-bandwidth radar altimeter signals, the large amount of input data requires a large number of training parameters when utilizing fully-connected neural network autoencoder layers \cite{Logue2023}.

In this work, we handle these challenges in two distinct ways. First, we take advantage of the inherent issues with reconstructing communication signals by focusing the autoencoder on reconstructing the well-defined radar signal and learning to remove the communication signal. This uniquely leverages the known limitation of these architectures for the benefit of our application space. Second, we leverage advances in autoencoder architectures by considering a Convolutional neural network (CNN) layer only architecture which allows for dimensionality reduction by making use of stride and pooling \cite{Nagar2022} and forgoes large fully connected layers like typical approaches consider. 

More specifically, this paper develops an end-to-end FMCW radar altimeter simulator used to train a CNN-based denoising autoencoder which directly performs interference mitigation on the received IQ signal return. The approach is demonstrated to be effective in removing several sources of narrowband interference from signals consisting of $40,000$ IQ samples in terms of improving SINR, RMS range estimation error, and false altitude reports generated by the radar altimeter.

The paper is outlined as follows. In Section \ref{sect:model} we present our radar altimeter simulation environment used to generate both the radar altimeter signals and interference signals, as well as to generate and estimate the radar return and ground estimate. In addition, this section presents the CNN-layer only autoencoder architecture. In Section \ref{sect:training}, we present the training and evaluation approach for the autoencoder using the developed simulation environment and in Section \ref{sect:results} present analysis of our proposed approach. Finally, in Section \ref{sect:conclusions} we present our conclusions and future work.

\section{Model Architecture / Methodology}\label{sect:model}

\subsection{Radar Altimeter Application and Dataset}

The autoencoder (AEC) model presented in this work is evaluated for its performance in a commercial aircraft altimeter system. Figure 1 depicts the structure of the radar altimeter simulation used to train and evaluate the AEC. To establish a realistic operating environment, the simulation incorporates a landing scenario with information regarding the flight trajectory and digital terrain elevation data (DTED). This terrain information is used to generate clutter, emulating the range response. 
 Continuous Wave Linear frequency modulation (CWLFM) signals are then transmitted through an AWGN channel while interference, either tone-based or QPSK is added. The signal processor block then takes in the received signals, produces a range profile, and makes range estimates whose error is evaluated in comparison to the true range.

\begin{figure*}[t]
    \centering
    \includegraphics[width=\textwidth]{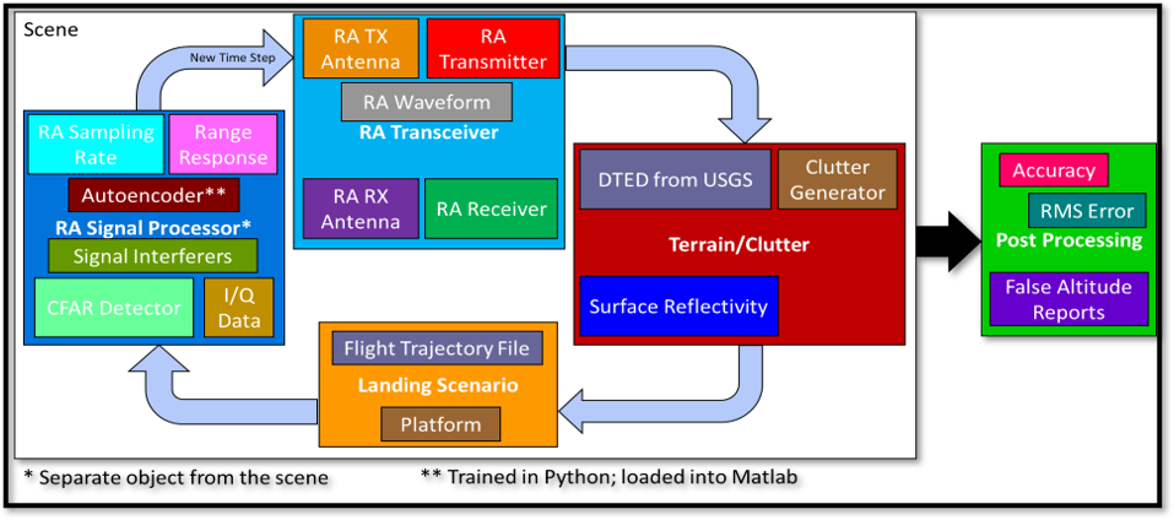}
    \caption{High-level block diagram of radar altimeter simulation used for both training data generation and evaluation of the trained model. Ideally, the latent representation of the FMCW signal consists of features unique to the signal of interest, such that unwanted interference can be discarded in the reconstruction process.}
    \label{fig:system_diagram}
\end{figure*}

\subsection{Autoencoder Model Architecture}

\begin{figure}[t]
    \centering
    \includegraphics[width=\columnwidth]{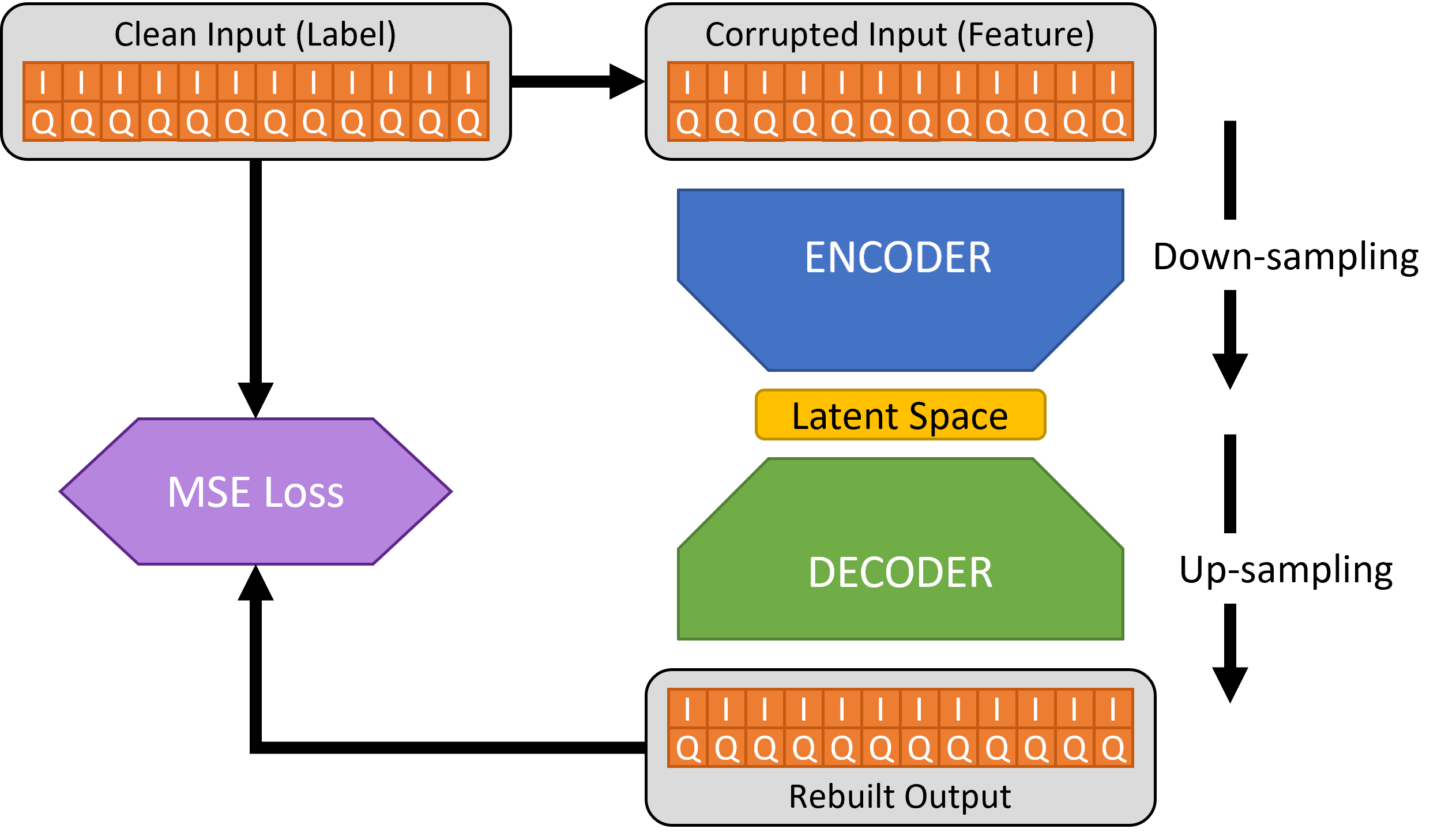}
    \caption{High-level block diagram of the denoising autoencoder used for interference mitigation.}
    \label{fig:autoencoder_highlevel}
\end{figure}

\begin{figure*}[t]
    \centering
    \includegraphics[width=\textwidth]{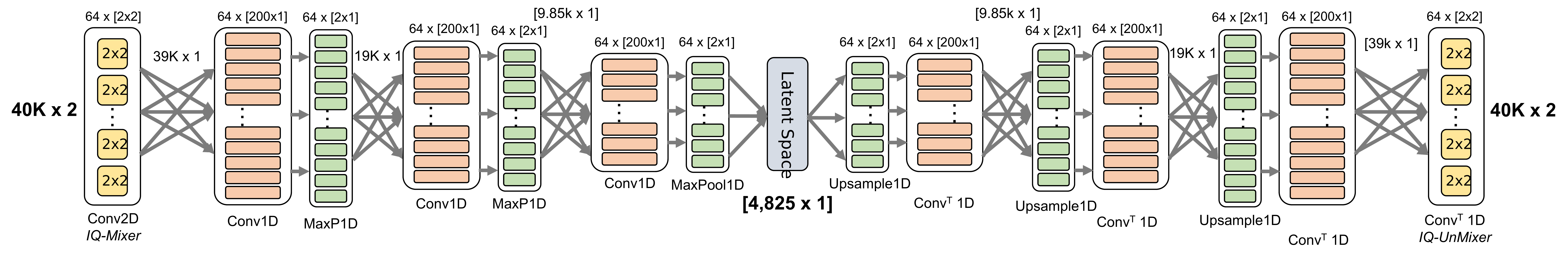}
    \caption{40k-Sample 30 MHz CNN Autoencoder Architecture}
    \label{fig:autoencoder_lowlevel}
\end{figure*}

The model architecture presented in this work uses an autoencoder structure in which the dimensionality of the input space is reduced to form a latent representation before expanding again to form an output space with the same dimension as the input space. This compression step forces the model to generalize patterns in the feature space while discarding most of the information. The autoencoder structure, often referred to as a bottleneck, is designed to perform a denoising operation on the inputs by removing unwanted information; i.e., nuisance parameters, and preserving the desired patterns in the feature space.
The application of this autoencoder to radio frequency (RF) signal processing introduces several non-traditional requirements such as the ability to handle complex-valued inputs and outputs [2]. There are multiple techniques for incorporating complex data into existing neural network architectures of which there are two main categories; those utilizing complex weights and biases, and those using only real weights and biases. The chosen technique for this model splits the real and imaginary parts into separate channels such that the input signals have dimension $\{N\times2\}$ and subsequently applies a 2D convolution operation with $\{2\times2\}$ kernels, reducing the signal to a 1-D sample vector. Although an alternative approach exists where the magnitude and phase of the signal are used instead of the real and imaginary components, the two-channel IQ format has been shown through this AEC architecture to be capable of adequately preserving the frequency/phase relationships of the input signal.

The denoising and interference mitigation capabilities of the AEC are made possible through the bottleneck structure which forces generalization on certain features and discarding of extraneous information, hopefully corresponding to the interference to be removed. The amount by which the dimensionality of the input signal is reduced in the latent representation is referred to as the compression ratio. Obtaining the optimal compression ratio is one of the main considerations driving the AEC design.

The primary distinction in the design of this AEC architecture lies in its use of solely CNN layers, setting it apart from more conventional designs utilizing some fully-connected layers. The CNN-only design is motivated primarily by the comparably smaller numbers of weight and bias parameters required to achieve a desired compression ratio given a certain input size. Fully-connected layers can achieve dimensionality reduction simply by reducing the size of subsequent layers, but as the number of parameters required for each layer is related to the product of its input and output dimension, large sample-length inputs become costly in terms of memory and computation. CNN layers achieve dimensionality reduction through the use of strided convolution or pooling layers, both of which serve as methods of down-sampling the signal. The parameter requirements for CNN layers are instead related to the kernel size, number of channels, and stride allowing for similar compression ratios for large signal sizes without ballooning memory and computational needs.

The AEC model architecture is depicted in Figure 3 highlighting the chosen structure and hyper-parameters. As mentioned, the model consists of only CNN layers with 4 CNN operations performed in the encoder and 4 transposed-CNN operations performed in the decoder. The first layer of the encoder and last layer of the decoder, called the ``IQ-Mixer'' and ``IQ-Unmixer'', are responsible for mixing the 2D IQ representation of the input signal down to a 1D real-valued output and then taking a 1D real-valued signal in the decoder and producing a 2D IQ output representation. Between the IQ-mixing layers there are 3 1D CNN-pooling pairs with max-pooling acting as the dimensionality reduction mechanism. The 3 1D CNN layers each have kernel sizes of 200 meaning that 64 unique kernels of size 200 are convolved with each layer.

In the encoder each CNN layer is followed by a max-pooling layer, and in the decoder each transposed CNN layer is preceded by a max-unpooling layer. This design behaves as a filter then down-sample operation in the encoder and an upsample then filter operation in the decoder. The choice of this overall CNN structure stems from the temporal correlation properties of the RF signal features and the ability for CNN-pooling pairs to mimic the behavior of classical filtering techniques such as adaptive filters but with added flexibility. With pooling-kernel sizes of 2 for each max-pooling and upsampling layer, the input dimensionality is reduced resulting in a compression factor of $\approx 16:1$, the refinement of which is discussed in the sequel.

The loss function employed for the AEC compares the rebuilt inferred output signals to the clean label signals using $L_2$ loss or MSE. The loss function compares the raw signals which allows for optimization over the full signal dimension.

\section{Training Approach}\label{sect:training}
The training approach for the CNN AEC model was geared towards the goal of signal shape generalization, meaning that the model should ideally preserve those features which can accurately reconstruct the CWLFM signal from a corrupted, interfered state. This training procedure was primarily an iterative process wherein different architectures are tested under the same dataset and evaluated for their generalization performance and range estimation capability.

The most challenging aspect of this process was the difficulty in training on long signals; i.e., those with 10k samples or more. Therefore, much of the initial testing was done on shorter-length signals of ~1000 samples with the hope of achieving similar results with larger length features after scaling up the kernel size. With this constraint, the initial refinement of the CNN architecture utilized 10,000-example datasets with ~1000 samples per signal feature.

Dataset generation was key to the refinement of the AEC architecture, allowing generalization of the overall waveform shape while avoiding over-fitting and under-fitting. The dataset was designed to expose the model to a representative sample of intended features as well as nuisance features such as corruption, noise, and interference. The ultimate target feature which the model was trained to preserve was the time-delay applied to the signals mimicking the path delay of the radar altimeter and containing the information needed to estimate the range through stretch processing. Signals were therefore given random time delays uniformly distributed within the range $[0, 0.01]$ proportional to the signal length, corresponding to true range values in the range $[0, 2000]$ meters. These delayed signals were then set aside as the clean signals, and dirty signals were produced by corrupting the clean signals with amplitude variation, adding white Gaussian noise (WGN), and adding interference of choice either consisting of tones or QPSK.

The amplitude variation was applied to vary the envelope of the CWLFM signals in a manner similar to the effect of constructive and destructive interference from a multipath fading channel. The varying envelop was generated as a band-limited Gaussian process having 10\% the bandwidth of the CWLFM signal and a standard deviation of 0.3. The Gaussian noise was then added to set the SNR level for the radar altimeter in the range $[-25, 30]$ dB. Interference was added through some combination of tones and QPSK signals.  Tones were spaced uniformly in frequency across the CWLFM bandwidth and having SIR values in the range $[-20, 20]$ dB. The QPSK interference was meant to mimic communications signals spanning some portion of the radar bandwidth and pulse duration, with bandwidths and durations uniformly spaced across frequency and time and SIR values in the range $[-20, 0]$ dB.

The process employed for model refinement and hyper-parameter tuning involved training the model on smaller 1000-length signal datasets of 10,000 examples and observing the performance in terms of rebuilt signal quality, range profile peak-to-sidelobe ratios, and range estimation accuracy. Figure 4 illustrates the process used for evaluation of the model wherein inference was performed on an evaluation dataset created in similar fashion to the training dataset. Stretch processing is performed on both the inputs and outputs of the AEC using a pre-transmission reference signal, similar to the operation of the actual altimeter system. The stretch-processed signals contain the same information content as the original signals, but this information is transformed to reflect the range estimation capability. The range profiles feature peaks in both positive and negative frequency due to the upsweep and downsweep portions of the CWLFM. The range estimate is obtained by detecting and averaging these two peaks. With this evaluation scheme in place, each trained model's effectiveness was determined by the peak-to-sidelobe ratios for its range profiles as well as the accuracy of its range estimates compared to the dirty signals.
\begin{figure}[t]
    \centering
    \includegraphics[width=\columnwidth]{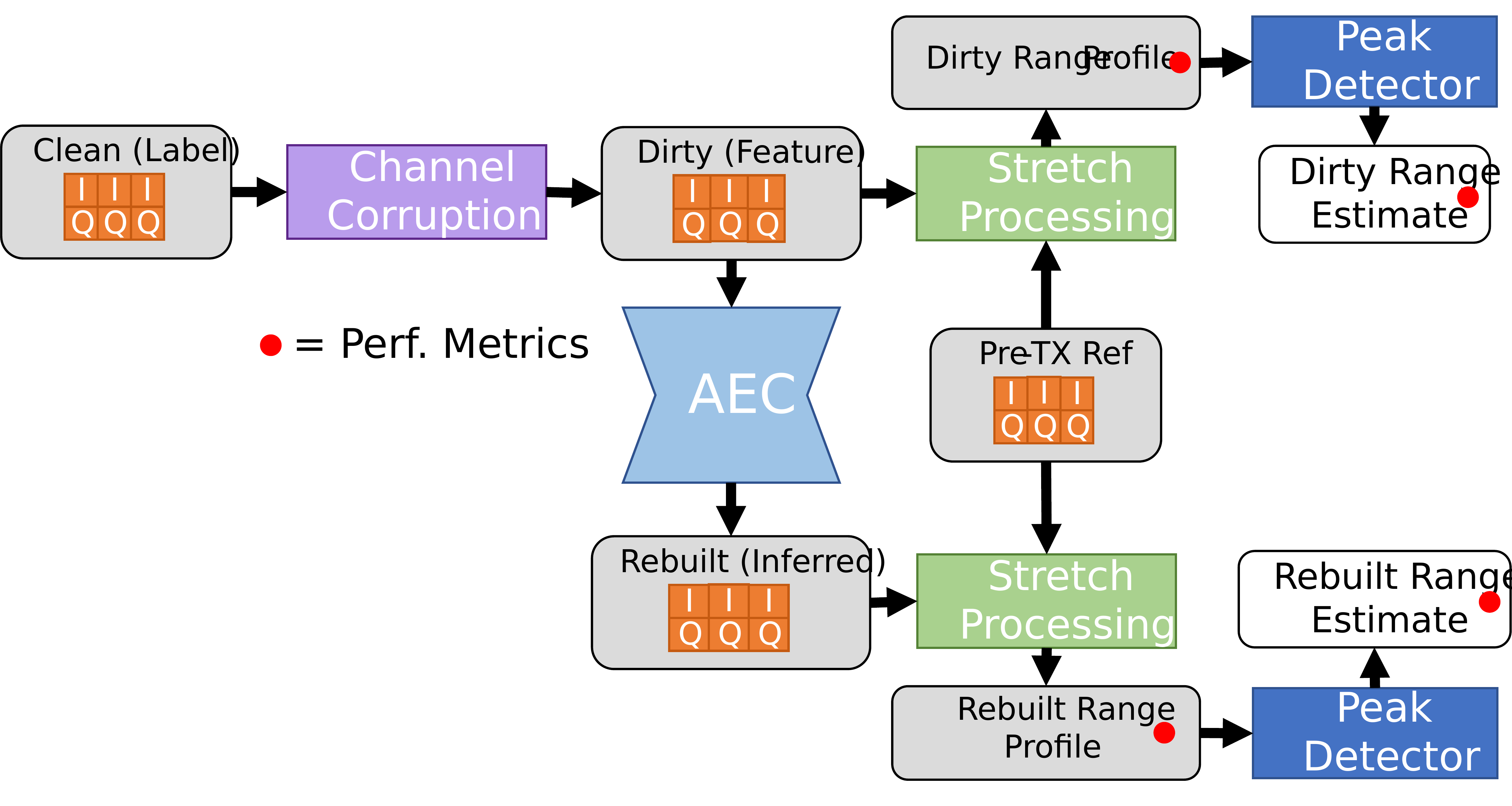}
    \caption{AEC Model Range Profile Evaluation Scheme}
    \label{fig:model_evaluation_diagram}
\end{figure}

The main hyper-parameters that were refined using this training and evaluation scheme were the CNN kernel size and the compression ratio of the AEC. It is clear that the kernel size chosen within the CNN layers has a significant impact on the model's ability to generalize on the signal shape. The authors of [4] pose the idea that the signal pattern elements such as symbols for communication signals should largely dictate the kernel size. Using the 1000-sample signals it was observed that a kernel size of about 300 produced the best results in terms of MSE as smaller kernels resulted in reduced pattern recognition and larger ones resulted in over-fitting behavior. As expected, larger input sizes have larger optimal kernel sizes, reflecting the need for longer temporal correlation to capture the signal shape in these cases.

Taking a similar approach towards optimizing the compression ratio, it was observed that latent space dimensions of between 10\% and 20\% of the input dimension produced the best results in terms of MSE. The existence of an optimal compression ratio for AEC architectures is supported by the idea that there is some optimal level of information retention which adequately represents the desired features while minimizing retention of nuisance parameters.

Taking these findings into account, the final model architecture presented prioritizes achieving the optimal compression ratio over the optimal kernel size. Kernel sizes of 200 are used for each layer despite the input dimension of 40k samples, a decision largely influenced by the computational cost associated with larger kernel sizes.

\section{Inference Results}\label{sect:results}
The AEC performance was evaluated based on its ability to improve range estimation performance under different interference conditions. To obtain a comprehensive evaluation of the AEC performance, inference tests were performed using both the AEC model evaluation scheme shown in Figure 4 and the end-to-end radar altimeter simulation. Two metrics were used for determining the quality of the range profile with the first being the peak-to-sidelobe ratios which provided a metric for interference mitigation capability and the second being the range estimation error. Together these metrics demonstrate the degree to which unwanted interference can be suppressed and the ability for the AEC model to improve range estimation accuracy when ideal pulse compression is employed.

\begin{figure*}[t]
    \centering
    \includegraphics[scale=0.45]{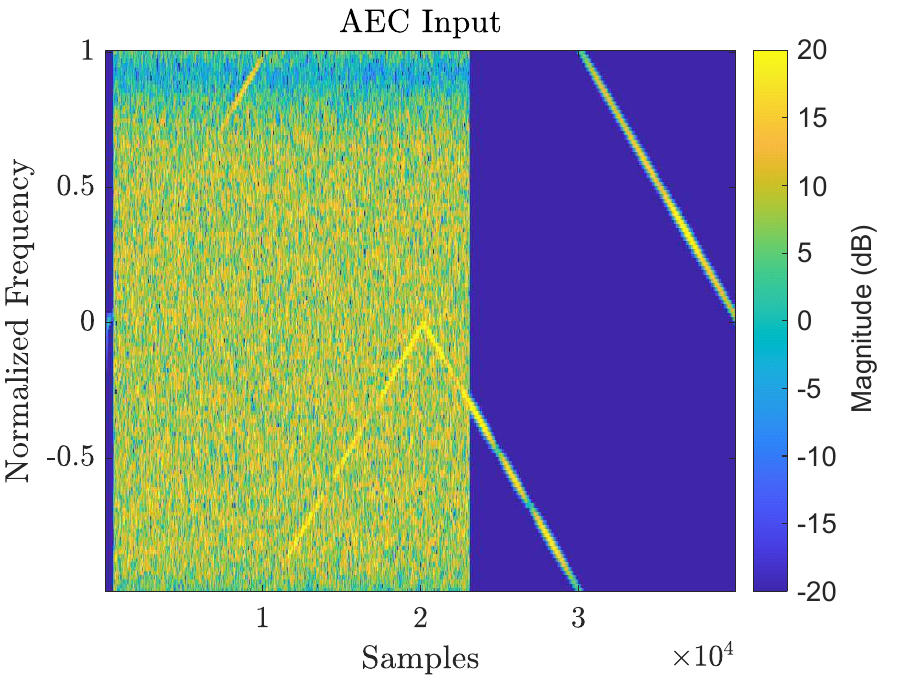}
    \includegraphics[scale=0.45]{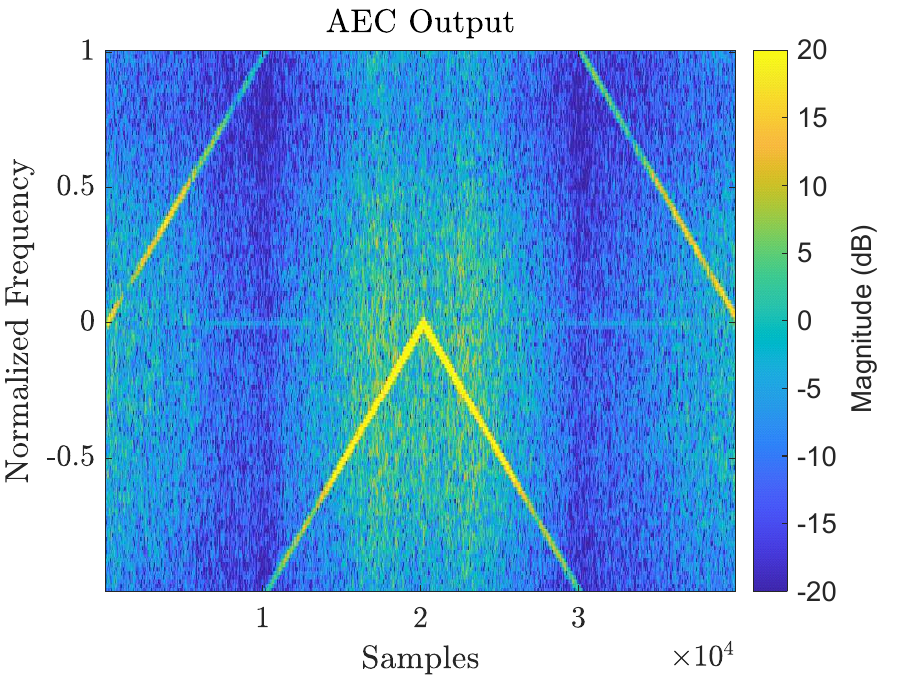}
    \caption{Example spectrograms of the received radar signal at the input and output of the autoencoder with QPSK interference.}
    \label{fig:QPSKinference}
\end{figure*}

\begin{figure*}
    \centering
    \includegraphics[scale=0.45]{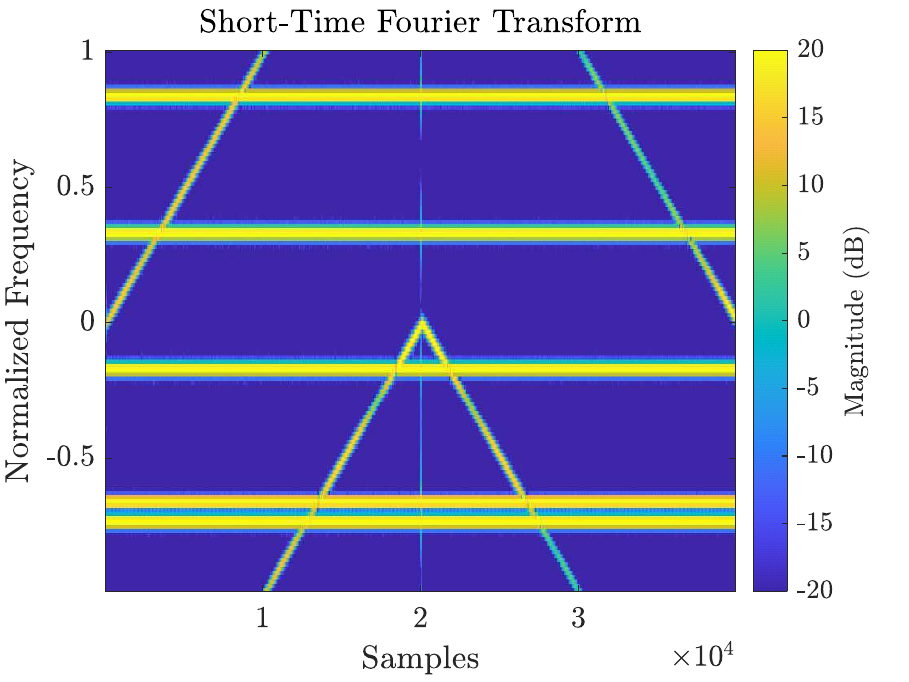}
    \includegraphics[scale=0.45]{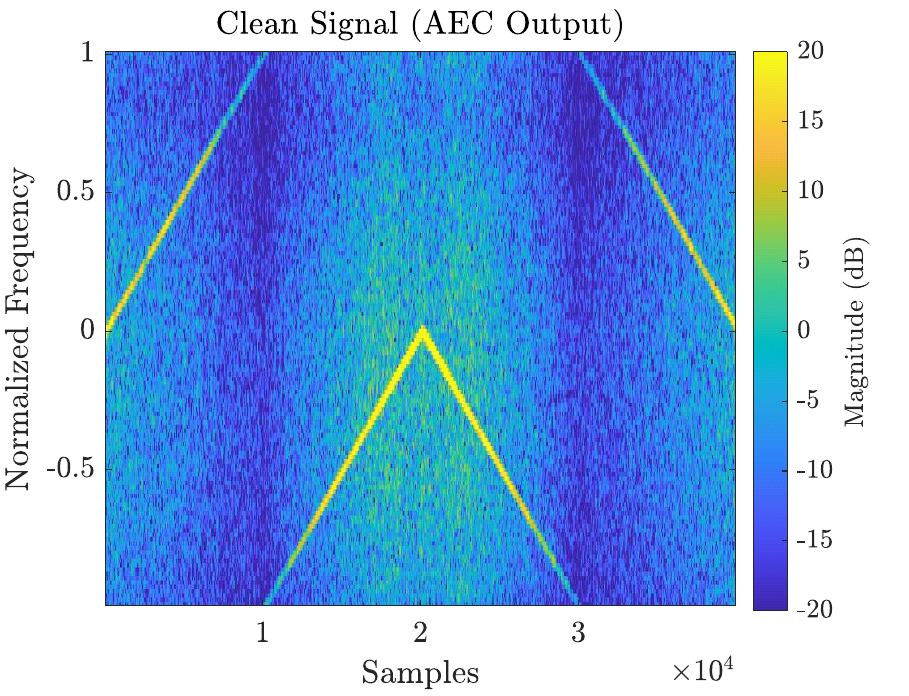}
    \caption{Example spectrograms of the received radar signal at the input and output of the autoencoder with 5 narrowband tone interferers.}
    \label{fig:toneinference}
\end{figure*}

An example of the qualitative AEC performance under harsh interference conditions is shown in Figure \ref{fig:QPSKinference}. The radar signal was propagated through the radar altimeter simulation, and was also exposed to additive co-channel QPSK interference at an SIR of approximately 0dB. We observe that the autoencoder model is able to effectively reconstruct the FMCW signal, even when the QPSK interferer occupies the entire signal bandwidth. 

An additional example is seen in Figure \ref{fig:toneinference}, where 5 narrowband interferers are are added to the received radar signal at an SIR of approximately -10dB. We observe that once again, the autoencoder is able to effectively reconstruct the desired FMCW radar signal


\begin{figure}[t]
    \centering
    \includegraphics[width=\columnwidth]{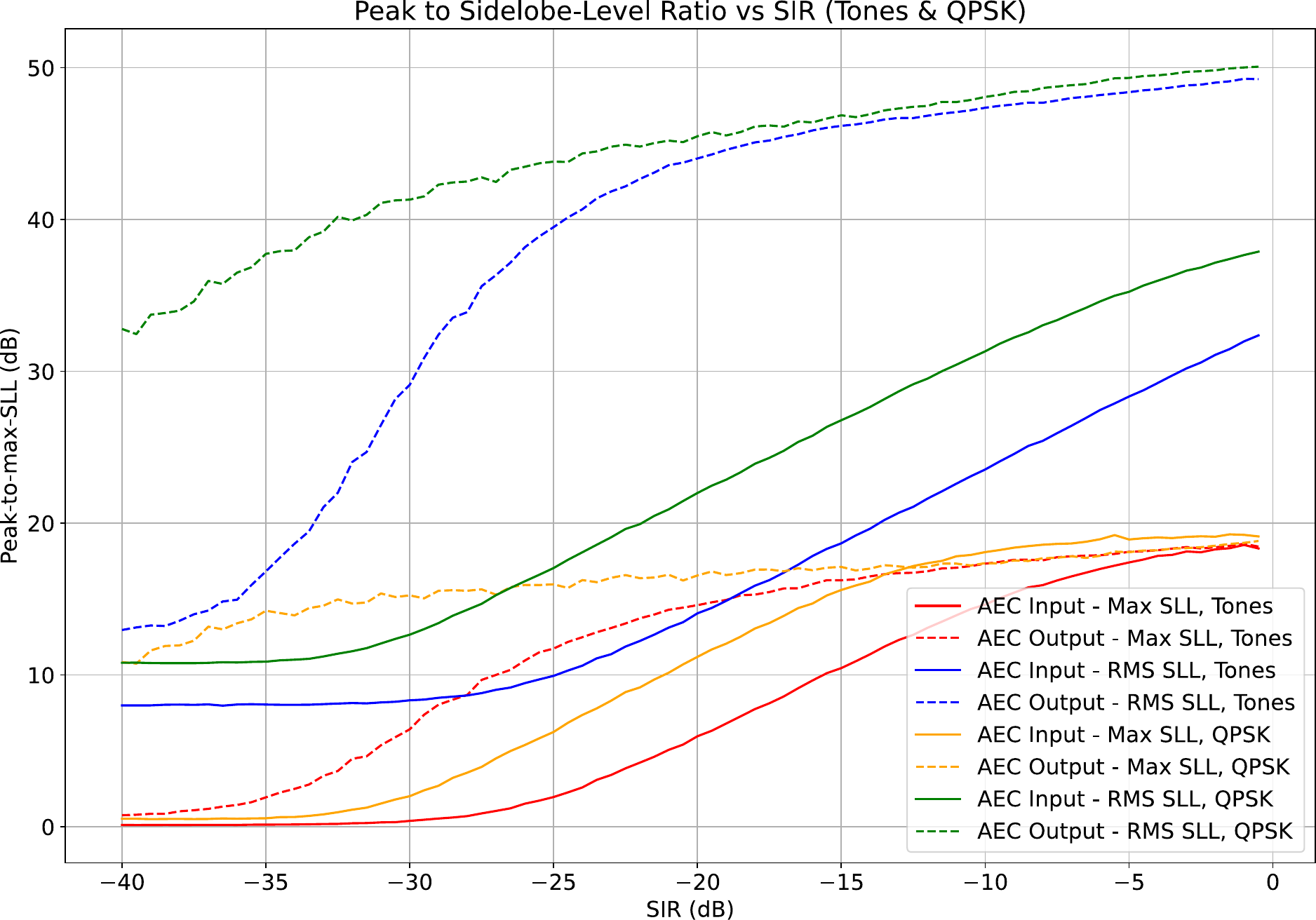}
    \caption{Range Profile Peak to Sidelobe-Level Ratios (Tones \& QPSK)}
    \label{fig:range_profile}
\end{figure}

As can be seen from Figures \ref{fig:QPSKinference} and \ref{fig:toneinference}, the AEC model shows significant improvement in the range profile quality when subjected to both tones and QPSK interference. The evaluation dataset used for this result consists of 1,000 signals each consisting of 40k- IQ samples with an SNR level of 0 dB and either tones or QPSK interference.

The AEC performance is evaluated using several metrics and is tested with stand-alone generation as well as the end-to-end radar altimeter performance when subjected to tone interference even 100 times more powerful than the CWLFM signal as in this case. QPSK interference poses a greater challenge to the AEC, though, due to its wide-band characteristics. Despite this, the AEC is still able to preserve the original signal shape.

\begin{figure}[t]
    \centering
    \includegraphics[width=\columnwidth]{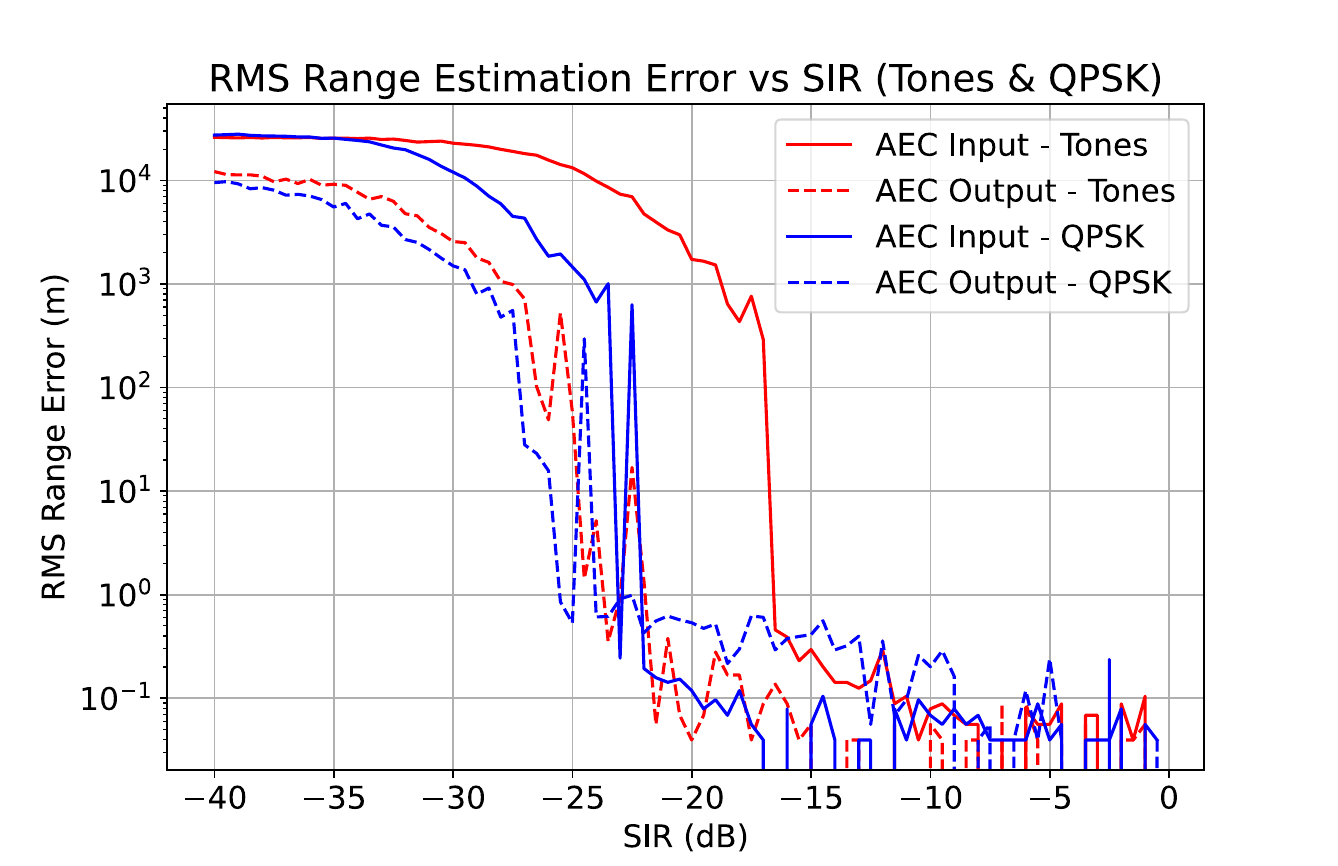}
    \caption{RMS Range Estimation Error vs SIR (Tones \& QPSK)}
    \label{fig:rms_range_error}
\end{figure}

The two primary metrics for AEC performance used here are the peak-to-sidelobe ratio in the range profile, which is shown in Figure \ref{fig:range_profile}, and the RMS range estimation error, which is shown in Figure \ref{fig:rms_range_error}. Figure \ref{fig:range_profile} shows how well the model is able to improve the quality of the range profile over varying input SIR values for both tones and QPSK interference. The model provides at best a 30 dB improvement in the range profile for -25 dB SIR tone interference. A similar result is obtained for -30 dB SIR QPSK.

Figure \ref{fig:rms_range_error} provides a direct indication of the range estimation improvement seen after applying the AEC. A threshold is observed where the range detector is no longer able to obtain a useful result, and this threshold is lowered once the AEC is applied in the case of tones and QPSK interference. The improvement for tone interference is notably higher than for QPSK interference.

\begin{figure}[t]
\centering
\begin{subfigure}{.5\columnwidth}
    \centering
    \includegraphics[scale=0.26]{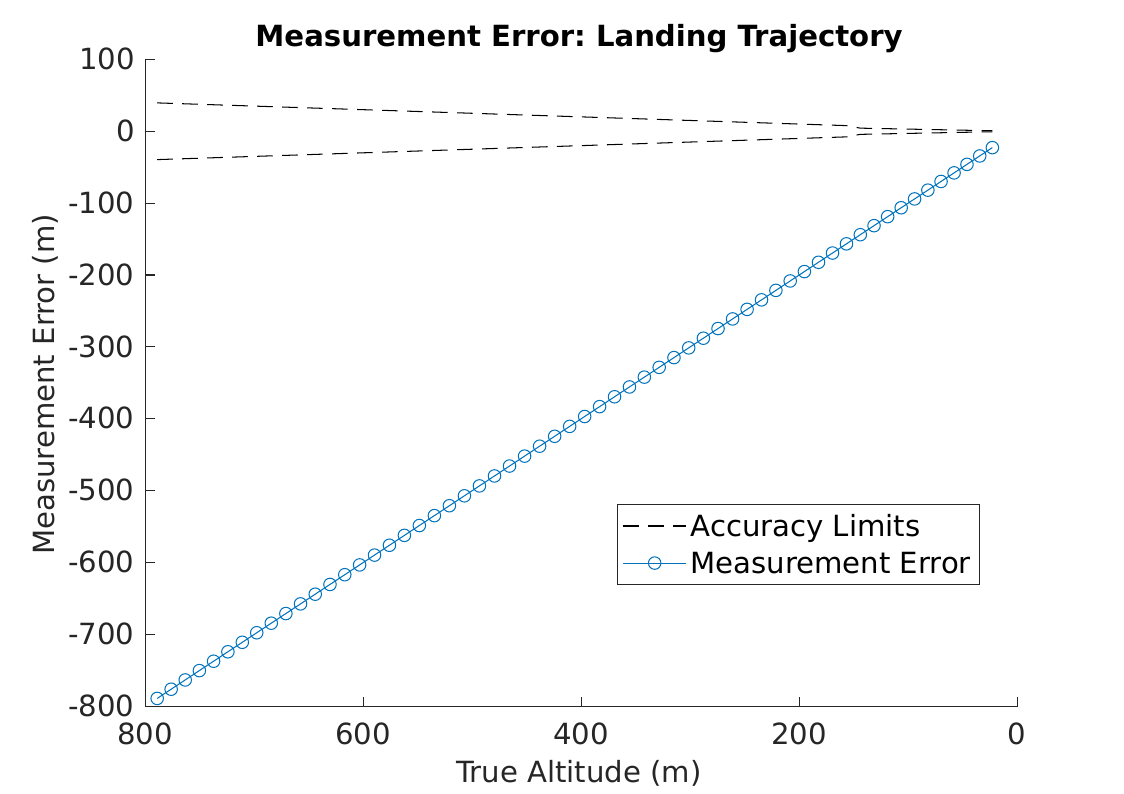}
    \caption{Range Error (No AEC)}
\end{subfigure}%
\begin{subfigure}{.5\columnwidth}
    \centering
    \includegraphics[scale=0.26]{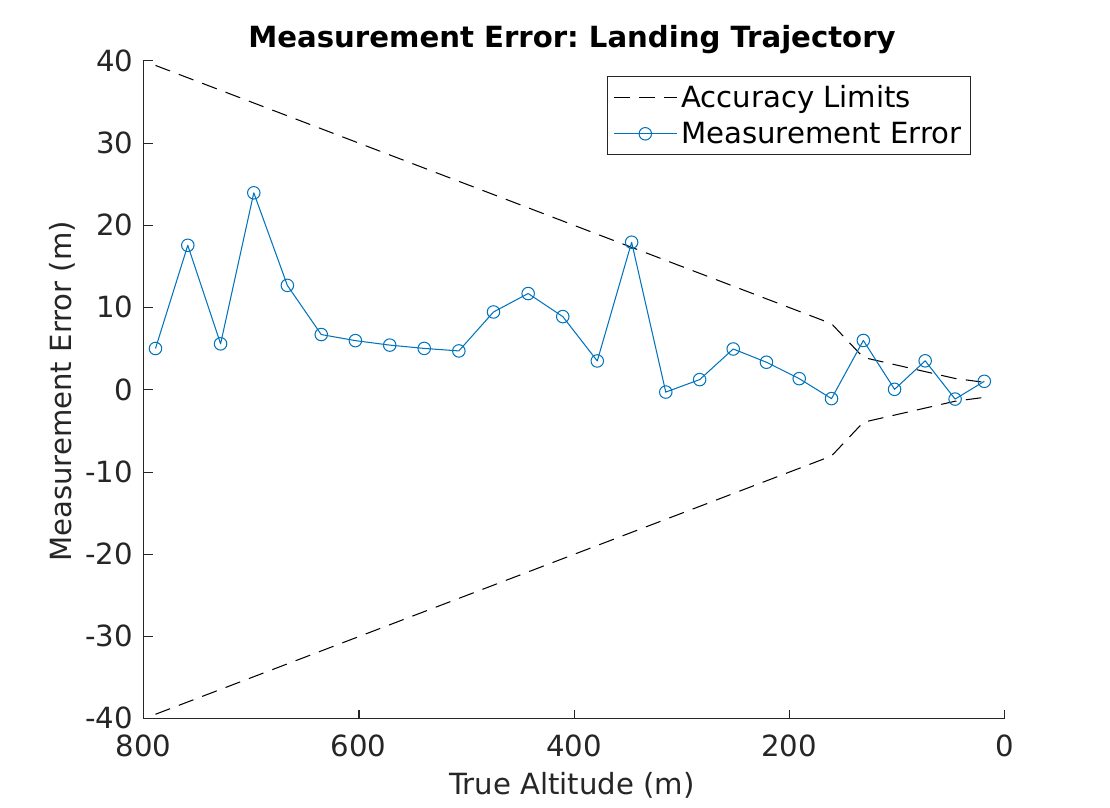}
    \caption{Range Error (With AEC)}
\end{subfigure}
\begin{subfigure}{.5\columnwidth}
    \centering
    \includegraphics[scale=0.26]{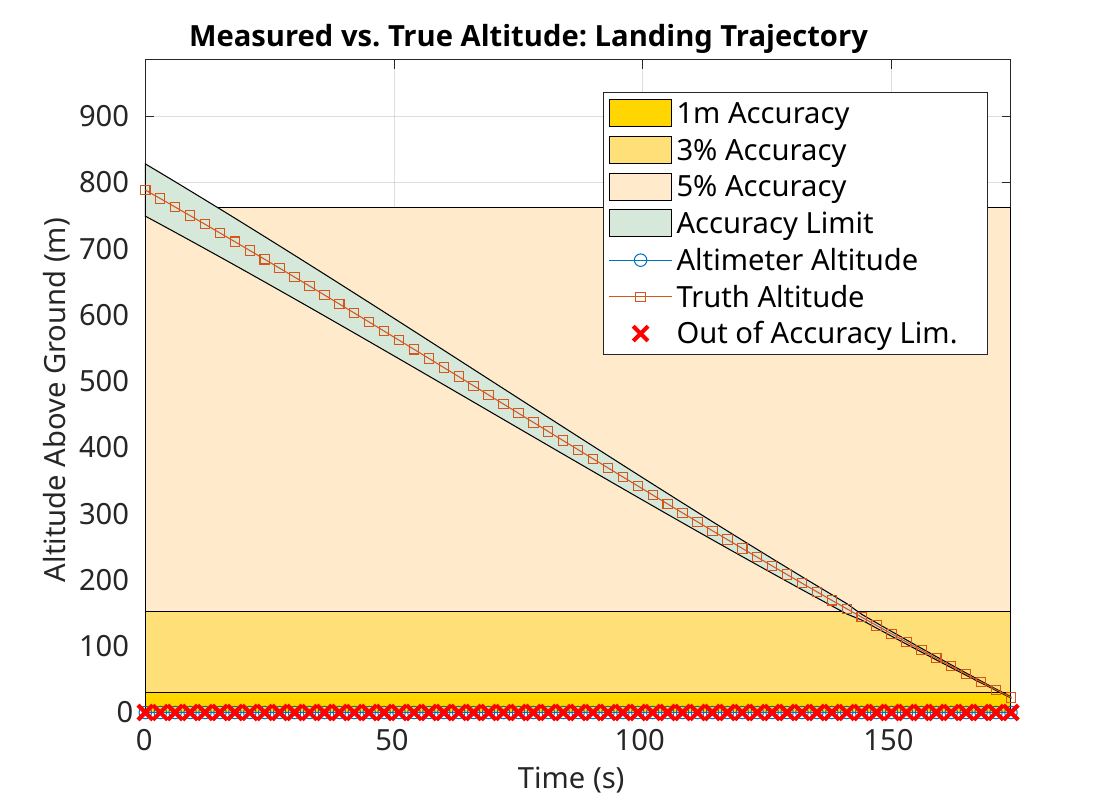}
    \caption{Landing Acc. (No AEC)}
\end{subfigure}%
\begin{subfigure}{.5\columnwidth}
    \centering
    \includegraphics[scale=0.26]{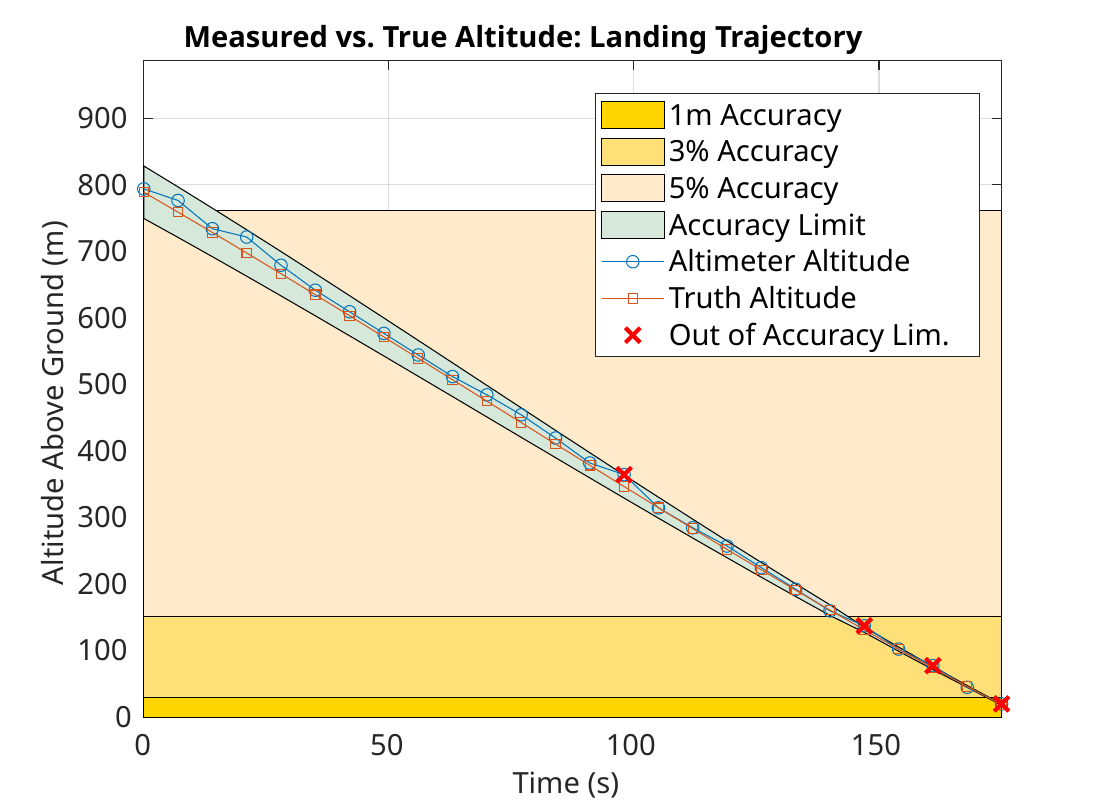}
    \caption{Landing Acc. (With AEC)}
\end{subfigure}
\caption{AEC Performance in End-to-End Radar Altimeter Simulation}
\label{fig:sim}
\end{figure}

Figure \ref{fig:sim} shows the end-to-end simulation performance of the AEC where the trained model is placed into the MATLAB-based altimeter simulation. The simulation runs through $175$s in a simulated aircraft landing trajectory, adds interference in the form of 5 tones each having -20 dB SIR and QPSK at -20 dB SIR, and calculates the estimated range using the signal processing block. 

\section{Conclusions and Future Work}\label{sect:conclusions}
We have presented a CNN-based autoencoder for interference mitigation in an end-to-end FMCW radar altimeter simulation. A lightweight model, capable of removing several sources of interference from received signals consisting of $40$K IQ samples was developed and tested on the simulated radar altimeter data. The interference mitigation capabilities of the autoencoder were evaluated in several respects, namely peak-to-sidelobe ratio, range estimation error, and number of false altitude reports in the altimeter simulation.

Several areas remain open for future work. From a practical perspective, the implementation should be further developed to facilitate real-time data processing. The model should be trained and validated on captured data, and generalization performance should be further evaluated.

Additionally, the interference mitigation capabilities of the present model could be compared to an array of conventional interference mitigation techniques, such as frequency diversity, the use of narrow main beams, and sidelobe null steering. The integration of waveform agility techniques at the transmitter \cite{thornton2020deep}, working in tandem with the denoising autoencoder could provide a more holistic interference mitigation solution.

\bibliographystyle{IEEEtran}
\bibliography{aecBib.bib}{}

\end{document}